# Spectroscopic Coronal Observations during the Total Solar Eclipse of 11 July 2010

**A.G. Voulgaris • P.S. Gaintatzis • J.H. Seiradakis • J.M. Pasachoff • T.E. Economou**

A.G. Voulgaris

*Evrivias 5c, GR-542 50 Thessaloniki, Greece*

P.S. Gaintatzis, J.H. Seiradakis (✉)

*Department of Physics, Section of Astrophysics, Astronomy and Mechanics, Aristotle University of Thessaloniki, GR-541 24 Thessaloniki, Greece*

e-mail: jhs@astro.auth.gr

J.M. Pasachoff (✉)

*Williams College—Hopkins Observatory, Williamstown, Massachusetts 01267, USA*
e-mail: eclipse@williams.edu

T.E. Economou

*Laboratory for Astrophysics and Space Research, Enrico Fermi Institute, University of Chicago, Chicago, IL 60637, USA*

**Abstract** The flash spectra of the solar chromosphere and corona were measured with a slitless spectrograph before, after, and during the totality of the solar eclipse of 11 July 2010, at Easter Island, Chile. This eclipse took place at the beginning of the Solar Cycle 24, after an extended minimum of solar activity. The spectra taken during the eclipse show a different intensity ratio of the red and green coronal lines compared with those taken during the total solar eclipse of 1 August 2008, which took place toward the end of Solar Cycle 23. The characteristic coronal emission line of forbidden Fe XIV (5303 Å) was observed on the east and west solar limbs in four areas relatively symmetrically located with respect to the solar rotation axis. Subtraction of the continuum flash-spectrum background led to the identification of several extremely weak emission lines, including forbidden Ca XV (5694 Å), which is normally detected only in regions of very high excitation, *e.g.*, during flares or above large sunspots. The height of the chromosphere was measured spectrophotometrically, using spectral lines from light elements and compared with the equivalent height of the lower chromosphere measured using spectral lines from heavy elements.

**Keywords** Eclipses • Corona • Chromosphere's Height • Continuum Background • Ionized Iron • Lower Chromosphere • Continuum Subtraction

## 1 Introduction

We continue our joint work on the solar chromosphere and corona, which started with the total solar eclipse of 29 March 2006 (Voulgaris *et al.*, 2010). On 11 July 2010, we observed the total solar eclipse at Easter Island in the South Pacific. The flash spectrum of the solar corona

A. G. Voulgaris *et al.*

was measured with a slitless solar spectrograph made by one of us (AV). Although Solar Cycle 24 had just begun, the emission lines from forbidden both Fe X and Fe XIV were clearly visible in the coronal spectra, whereas only the cooler of the two lines were detectable during solar minimum. This change in ratio indicates an increase in the temperature of the corona as solar activity is on its way to the next solar maximum, perhaps the last solar-activity cycle for some decades (Pasachoff, 2011; Pasachoff and MacRobert, 2011).

Throughout the recent solar minimum, the total absence of energetic flares as well as the general low solar activity resulted in a drop of the corona temperature. During that time, significant changes were observed in the intensities of the forbidden Fe X and Fe XIV emission lines, corresponding to ionization temperatures of $1.2 \times 10^6$ K and $1.8 \times 10^6$ K, respectively. Other strong lines are available in the infrared, as discussed by Pasachoff, Sandford, and Keller (1978) and, with improved techniques, by Habbal *et al.* (2011). During the solar minimum, many known multiply ionized coronal emission lines are not detectable, since they require a much higher temperature. An example of such elements is forbidden Ca XV (5694 Å), with an ionization temperature of $2.3 \times 10^6$ K, whose emission lines have never been reported during solar minimum. See also Wagner and House (1968).

Studies of the total solar eclipses from the ground are part of an effort to understand the processes that are going on in the solar corona with ever improving ground equipment (Singh *et al.*, 2011) and supplementing the results obtained from space missions (Pasachoff, 2009a, 2009b).

From the work of Mazzotta *et al.* (1998), it can be concluded that the multiply ionized elements with high ionization potential continue to emit even at lower temperatures, albeit with extremely low signal, near or at the limits of the continuum. This extremely faint signal from the above-mentioned elements explains why such lines are not directly observed during the solar minimum. Faint emission lines are barely detected in the spectra.

In this article, we concentrate on the analysis of flash spectra that were obtained during the 2010 total solar eclipse and the possibility of detecting extremely low intensity emission lines from multiply ionized elements in the solar corona. These extremely faint emission lines are easily recognized, since they follow the circular solar limb pattern, similar to the stronger emission lines.

**2. Observations and Data Recording**

2.1. The Total Solar Eclipse of 11 July 2010, at Easter Island

We observed the total solar eclipse from Easter Island, Chile, South Pacific (longitude $-109^\circ$ 25´ 29.2´´ and latitude $-27^\circ$ 08´ 36.3´´) under exceptionally clear weather conditions, in spite of the general weather variability of the previous days and rain until the very last morning of the day of the eclipse. The observations took place at the premises of the Vai Moana Hotel, 2

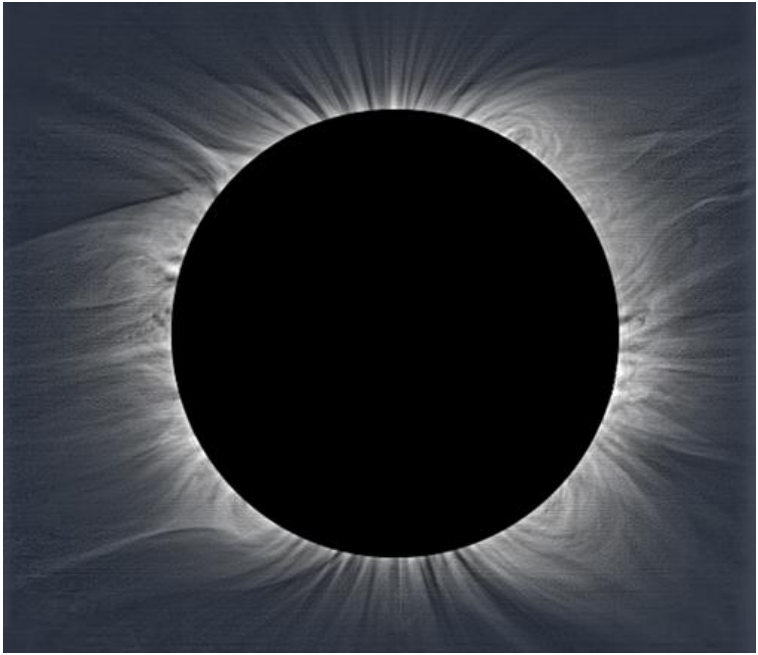

**Figure 1** The inner and middle corona of the total solar eclipse of 11 July 2010. Twenty-five images, of several exposure times, have been composed using a phase-correlation method. In the composed image, local adaptive filters have been applied for the enhancement of corona's structure. North is up.

Km from Hangaroa, the capital of the island. The observing site was 26.6 km away from the central line of the eclipse. The duration of the total eclipse at our site, uncorrected by the limb profile, was 4 minutes and 41.5 seconds; the Kaguya-LRO derived profile gave a duration of 4 minutes and 35.4 seconds. The middle of totality occurred at 20:10:49.8 UT (Espenak and Anderson, 2008; Jubier, 2010) with local time = UT −6 h, or 14:10. As is typical near solar minimum, the corona was extended equatorially without high-latitude streamers and with only plumes showing near the poles (Figure 1). The corona's structure in Figure 1 was enhanced using previously published digital techniques of Druckmüller (2006, 2009). Pasachoff *et al.* (2011a) provide extensive coverage of coronal imaging of this eclipse from the same site on Easter Island, from Tatakoto in French Polynesia, and from several solar spacecraft. The solar-activity cycle had risen considerably since the China/Marshall-Islands observations of the 22 July 2009 total solar eclipse, for which the sunspot number was essentially zero (Pasachoff *et al.*, 2011b).

### 2.2. The Slitless Spectrograph and Data Recording

A Canon 40D digital camera and a slitless spectrograph that have been described by Voulgaris *et al.* (2010) were used to record the flash spectrum. We used ISO 400 with exposure time 1/8000 second during the partial eclipse, between 1/5000 second and 1/1000 second, a few seconds before the second and after the third contact and between 1/50 second and 1/25 second during the total phase. The flash spectrum was captured before, during, and after the total eclipse in 179 images. In each image, the entire solar disk corresponded to 275 pixels or 5065 km pixel$^{-1}$. We show the dispersion of the spectrograph as a function of wavelength in Figure 2.

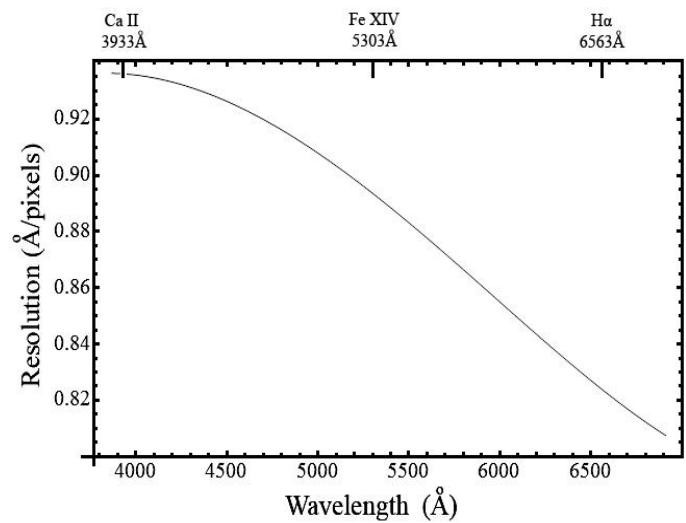

**Figure 2** The dispersion of the spectrograph as a function of wavelength.

## 3. Data Analysis and Image Processing

During totality, the Moon completely covers the photosphere and light comes only from the chromosphere and the solar corona. Light from the corona derives mainly from three sources: The continuum (*K*), the emission (*E*), and the Fraunhofer (*F*) coronas (*i.e.*, Golub and Pasachoff, 2010) and references in Pasachoff (2010). A number of telluric absorption lines are generated in Earth's atmosphere (mainly from $O_2$ and $H_2O$), whose intensities depend on the thickness of the atmosphere and on the terrestrial water vapor content. These lines are detectable in bands at the yellow ($H_2O$ absorption, 5875.6 Å – 5990.9 Å), orange ($O_2$ absorption, 6276.6 Å – 6310.6 Å, band-a), red ($H_2O$ absorption, 6472.4 Å – 6586.6 Å), and near infrared ($O_2$ absorption, 6867.1 Å – 6924.1 Å, band-B) part of the spectrum (Buil, 2011).

During totality, the flash spectrum is dominated by light from the *K*- and the *E*-corona. The *E*-corona in visible light is dominated by strong emission lines of forbidden Fe XIV (green), forbidden Fe X (red), and several other weak lines, which are difficult to detect as they are overwhelmed by strong continuum (K). The continuum background (*K*-corona) originates from scattered photospheric light and does not give any significant information of the energy content of the solar corona. So, we applied a set of techniques that resulted in significant reduction of the background continuum. Firstly, we combined a large number of images in order to increase the signal-to-noise ratio. Then the continuum was subtracted from this composite image, resulting in a clear view of the *E*-corona and the chromospheric emission lines. Finally, the emission lines were enhanced with a linear transfer function. In Figures 3 and 5 we assume that the flash spectrum extends along the horizontal axis (*x*). The range of pixel values in all gray images is normalized to double precision real numbers (64-bit) between 0 (black) and 1 (white).

Before proceeding with the analysis, we applied two corrections to the data.

*(i)* Because the dispersion of the spectrograph (Å pixel$^{-1}$) is not a linear function of the wavelength (Figure 2), the spectra were linearized with respect to wavelength.
*(ii)* Because we used a slitless spectrograph, all spectral lines follow the circular shape of the chromosphere. Furthermore, this circular shape is distorted to elliptical shapes of varying semimajor *x*-axes, as we move away from the optical axis of the spectrograph.

The relevant corrections were made using Wolfram's Mathematica v.8.0 by one of us (P.G.). The procedure used to make the above corrections is described in Section 3.3.

3.1. Alignment

In order to produce a final composed spectrum at third contact (Figure 3a) the available images (15 in total) had to be (*i*) aligned and (*ii*) added taking into account the exposure time of each image (Equation (2)). The alignment was achieved using the phase-correlation method (De Castro and C. Morandi, 1987; Reddy and Chatterji, 1996; Druckmüller, 2009). In particular, for the calculation of the translation ($\Delta x$, $\Delta y$) of an image $A(x,y)$ to an image $B(x,y)$ the following expression was used:

$$(\Delta x, \Delta y) = \arg\max_{(x,y)} \left[ F^{-1} \left( \left( \frac{F(A_{xy})\overline{F(B_{xy})}}{|F(A_{xy})+\varepsilon||\overline{F(B_{xy})}+\varepsilon|} \right) \exp\left(-\frac{u^2+v^2}{2\sigma^2}\right) \right)_{uv} \right]_{xy}, \quad (1)$$

where $F$ is the Fourier transform, $F^{-1}$ is the inverse Fourier transform, $\varepsilon$ is a sufficiently small non-zero real number, and $\sigma$ is the standard deviation of a Gaussian function. The quantities $(x,y)$ are the spatial coordinates and $(u, v)$ are the corresponding frequency coordinates. For the parameters $\varepsilon$ and $\sigma$ we used the values $\varepsilon = 0.0005$ and $\sigma = 2$. Using the above transformation, the 15 images of the flash spectrum that had been captured in the time interval ($t_{3rd} - 45\text{sec}, t_{3rd}$) (where $t_{3rd}$ is the instant of third contact) were aligned, with respect to the solar corona. Then, the aligned images were averaged:

$$I(x,y) = \frac{1}{n}\sum_{i=1}^{n}\frac{P_i(x,y)}{t_i}, \quad (2)$$

where $t_i$ is the exposure time of image $P_i(x,y)$ and $n = 15$ is the total number of images to be combined. The composed image $I(x,y)$ is presented in Figure 3a, while a cut across the y-axis (white line on either side of Figure 3a) is presented in Figure 4a.

3.2. Continuum Estimation and Subtraction

The continuum background intensity is not smooth along the y-axis; however, it extends smoothly along the x-axis, a fact that allows us to estimate it. Then, we can subtract it from the composed image $I(x,y)$. The continuum background can be estimated by applying a sequence of digital filters as presented below.
Firstly, we calculate the image $A_r(x,y)$:

$$A_r(x,y) = \min\{I(z,y): z \in [x-r, x+r]\}, \quad (3)$$

where $r$ is twice the maximum base width (in pixels) of the strongest emission lines in Figure 3a. This was found to be r = 50 pixels. Then, by using the convolution operator, we smooth the image $A_r(x,y)$ along the x-axis and we estimate the continuum background $I_c(x,y)$:

$$I_c(x,y) = A_r(x,y) * G(x) = \int_0^{x_{max}} A_r(t,y) G(x-t) dt, \quad (4)$$

where $G(x)$ is the Gaussian function with $\sigma = 50$ pixels. The image $I_c(x,y)$ is a very good approximation of the continuum background spectrum (Figure 3b and 4b). Finally, the difference

$$I_E(x,y) = I(x,y) - I_c(x,y) \quad (5)$$

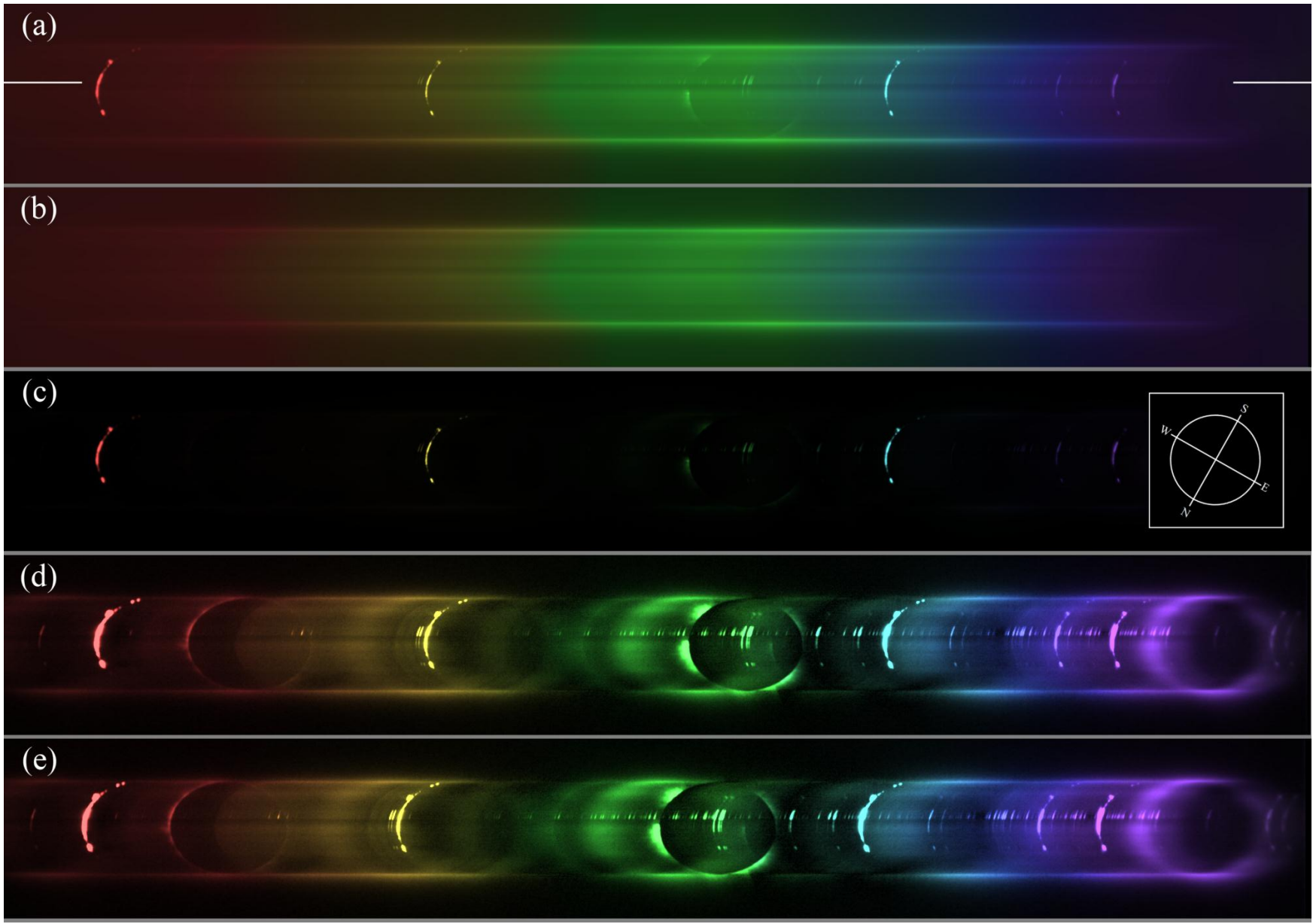

**Figure 3** From top to bottom: a) Composed image $I(x,y)$. b) Continuum background $I_c(x,y)$. The North -- South axis in all images is rotated by 29.5° clockwise with respect to the vertical axis. c) E-corona $I_E(x,y)$. d) E-corona $I_E(x,y)$ linearly enhanced. e) Linearization of the enhanced image with respect to wavelength $T_1 I_c(x,y)$. In this spectrum the resolution (Å pixel$^{-1}$) along the $x$-axis is constant.

Spectroscopic Coronal Observations during the Total Solar Eclipse

is the spectrum of the *E*-corona (Figure 3c, Figure 4c). In order to enhance the details of the displayed spectrum of the *E*-corona, a suitable linear transfer function was applied to the image $I_E(x,y)$ (Figures 3d and 4d).

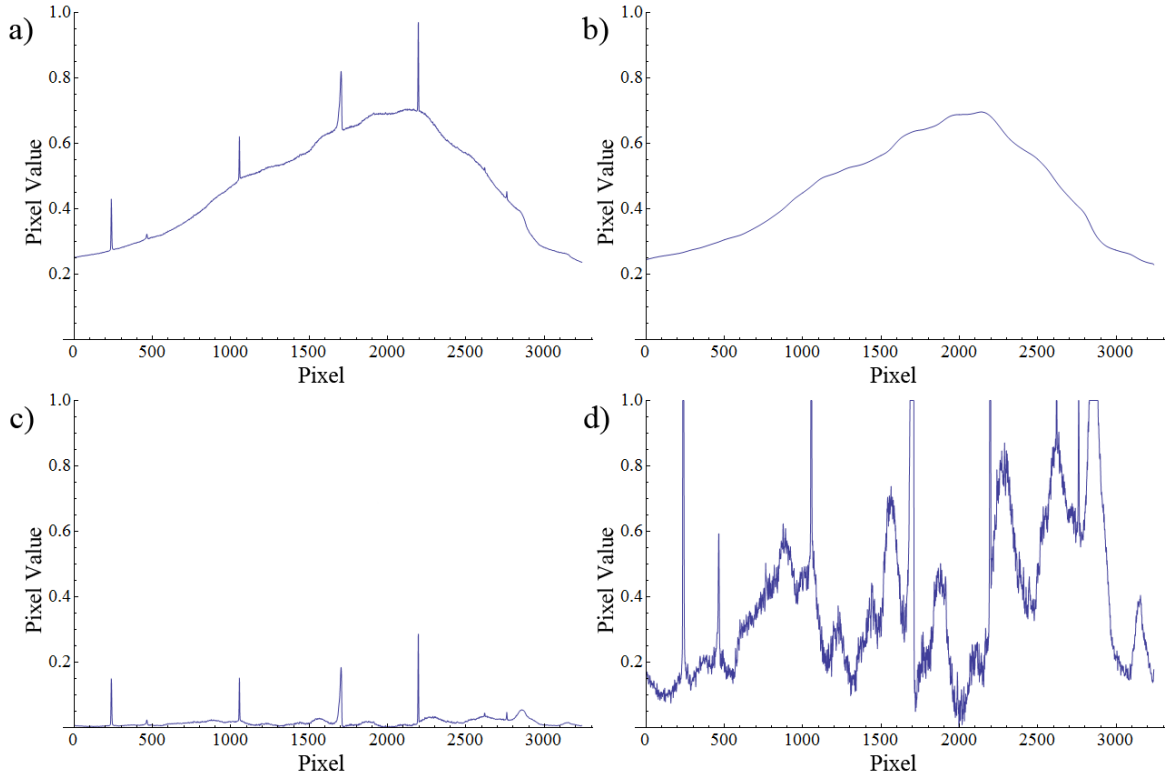

**Figure 4** a) Top left: the cut across the *y*-axis of Figure 3a. b) Top right: the cut across the *y*-axis of Figure 3b at the same *y*-location as the white line either side of Figure 3a. c) Bottom left: the cut across the *y*-axis of image $I_E(x,y)$ at the same *y*-location as the white line either side of Figure 3a. d) Bottom right: the cut across the *y*-axis of Figure 3d at the same *y*-location as the white line either side of Figure 3a.

### 3.3. Linearization of the Spectrum

It is obvious from Figure 2 that the dispersion (Å pixel$^{-1}$) of the spectrograph depends on the wavelength. This means that in order to proceed with the identification of emission lines we have to linearize our spectra. Let *w* be the width of $I_E(x,y)$ and $\lambda_{min}$ ($\lambda_{max}$) the smallest (largest) recorded wavelength in the right (left) limit of $I_E(x,y)$. Let us define a cut $(x,y_0)$ at an arbitrary height $y_0$ across the spectrum and let $\lambda(x)$ be the corresponding wavelength along the cut. Figure 2 shows exactly the absolute derivative of $\lambda(x)$ as a function of wavelength. The function $\lambda: [0,w] \rightarrow [\lambda_{min},\lambda_{max}]$ of $I_E(x,y)$ is strictly monotonic and can be estimated by fitting a smooth curve on the points $(x_i, \lambda_i)$ of emission lines of the flash spectrum that are easily recognized (*e.g.* H, He, Na, Mg, Ca II) of the flash spectrum. For our purpose, we used a fourth-order polynomial. The linearization of the spectra was achieved via the transformation

$$T_1(x, y) = \left( w \frac{\lambda(x) - \lambda_{max}}{\lambda_{min} - \lambda_{max}}, y \right), \qquad (6)$$

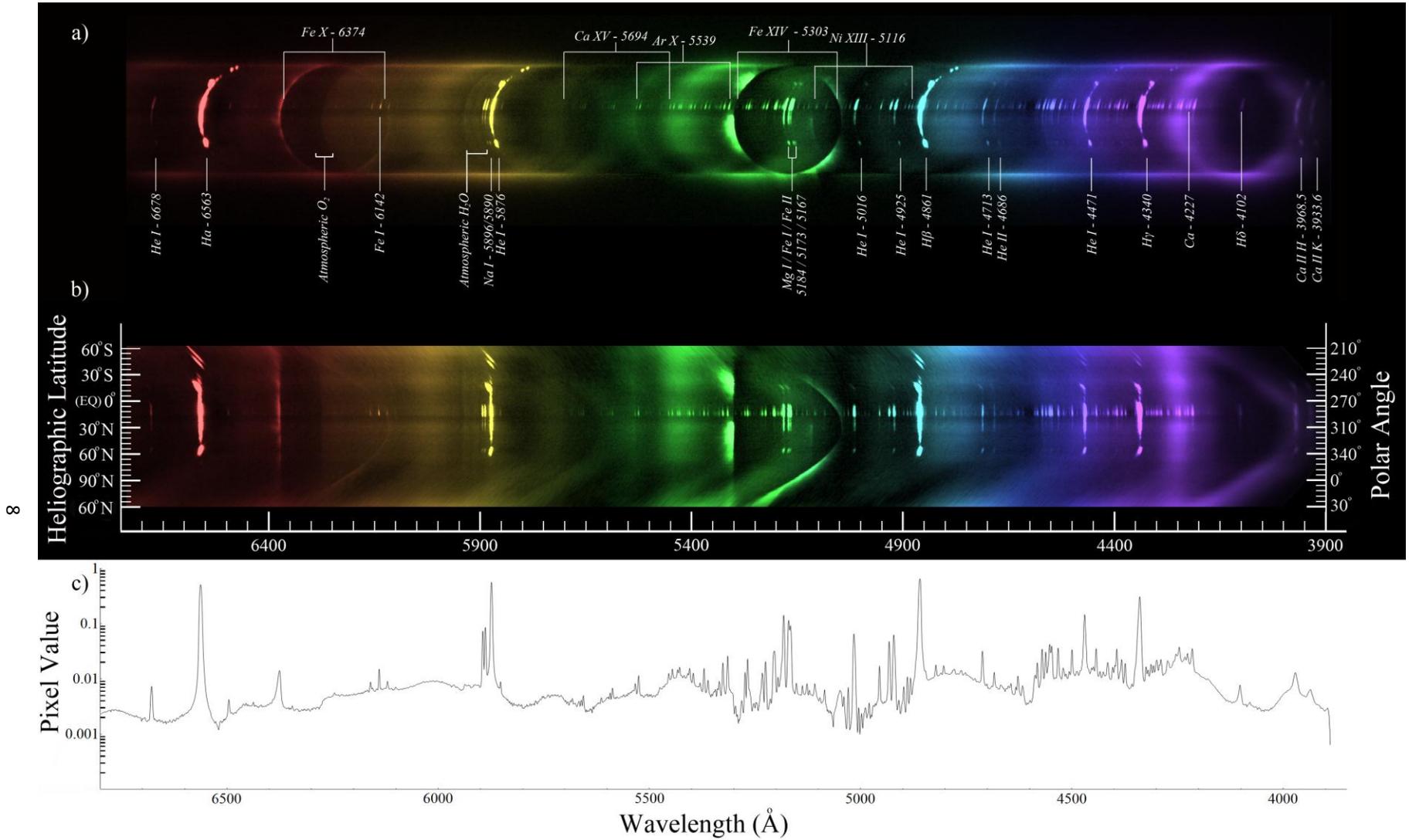

**Figure 5** From top to bottom: a) Image $T_2T_1I_E(x,y)$ linearly enhanced. b) Linearization with respect to circumference of the solar disk $T_3T_2T_1I_E(x,y)$ linearly enhanced. c) $f(\lambda)$ with $\varphi_1 = 270°$ and $\varphi_2 = 280°$.

where each pixel at position $(x,y)$ of the spectrum $I_E(x,y)$ is mapped to the position $T_1(x,y)$ of the new spectral domain $T_1I_E(x,y)$ (Figure 3e).

We notice that each circumference of the solar disk in the linearized domain (Figure 3e), as a function of wavelength, in the flash spectrum $T_1I_E(x,y)$, is approached by an ellipse, with minor axis [2b] parallel to the y-axis and major axis [2a] passing through the centers $(x´, y_0)$ of the solar images in the spectra. This means that the linear transformation $T_2(x,y) = (x, ay/b)$, where each pixel at the position $(x,y)$ of the spectrum $T_1I_E(x,y)$ is mapped to the position $T_2(x,y)$ of the new spectral domain $T_2T_1I_E(x,y)$, transforms the ellipses to circles of radius $a$ (Figure 5a).

Each pixel at position $(x,y)$ of the spectral image $T_2T_1I_E(x,y)$ lies on a western semicircle of radius $a$ centered at a point $(x´, y_0)$. If $\theta$ is the polar angle of this pixel, then $x = a \cos\theta + x´$ and $y = a \sin\theta + y_0$. For the linearization of the western semicircumference we used the transformation

$$T_3(x, y) = (x - a + a\cos\theta, y_0 + a\pi - a\theta) \tag{7}$$

where each pixel at position $(x,y)$ of the image $T_2T_1I_E(x,y)$ is mapped to the position $T_3(x,y)$ of the new image $T_3T_2T_1I_E(x,y)$ (Figure 5b). In image $T_3T_2T_1I_E(x,y)$ the horizontal x-axis is proportional to the wavelength and the vertical y-axis is proportional to the polar angle. Finally, the function

$$f(\lambda) = \frac{1}{\Delta\varphi} \int_{\varphi_1}^{\varphi_2} T_3T_2T_1I_E(\lambda, \varphi) \, d\varphi \tag{8}$$

is the integrated intensity of the emission lines along the circumference of the solar images in the interval of polar angles $[\varphi_1, \varphi_2] = [270°, 280°]$ (Figure 5d).

## 4. The Emission Spectrum of the Solar Chromosphere and Corona

4.1. Emission Lines of the Solar Chromosphere

The chromospheric emission lines Hα, Hβ, Hγ, Hδ, He I (D$_3$), Ca II H and K (Pasachoff and Suer, 2009; Voulgaris *et al.,* 2010) as well as the emission lines of the heavier elements of the lower chromosphere of Mg I, Fe I / II, Na I (D$_{1, 2}$) were visible just a few seconds after second contact and a few seconds before third contact (namely, after the photosphere was completely covered by the Moon) (Figure 5a). At the same time, on the eastern and western limbs, prominences in the emission lines from H I (Balmer series), He I, and from Mg I were also detected.

4.2. Emission Lines of the Solar Corona

Besides the chromospheric emission lines, the two basic coronal forbidden lines Fe X (6374 Å – ionization potential 235 eV) and Fe XIV (5303 Å – 355 eV) (Jefferies *et al.*, 1970; Gibson, 1973) were readily seen. In order to investigate the spatial distribution of the above lines we have magnified and rearranged part of Figure 5b (see Figure 6). It is obvious that Fe XIV emits at two separate medium heliographic latitudes (6° – 36° N and 24° – 66° S from the solar

Equator) and at extended heights above the chromosphere, where the temperature is higher, while there is no emission in the equatorial area (24° S – 6° N). On the contrary, Fe X emission is relatively homogeneous and covers an extended area along the chromosphere and its distribution occurs at low heights, where the temperature is lower. The regions covered by the two emission lines do not coincide. At this point, it should be mentioned that during the 29 March 2006 eclipse, which occurred well past the solar maximum, the distribution of forbidden Fe XIV was much closer to the solar Equator (– 20° to +15°) (Voulgaris *et al.*, 2010).

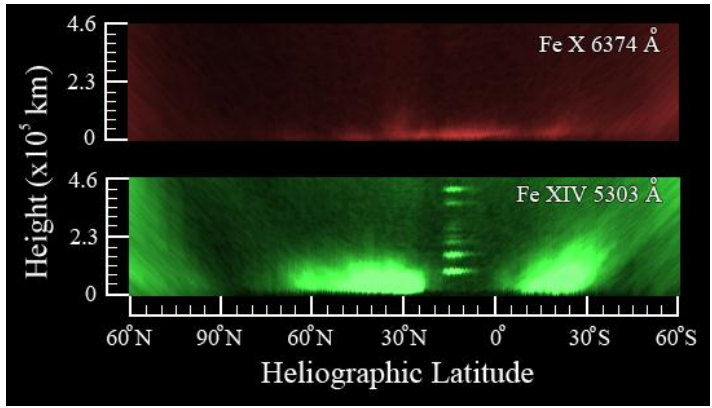

**Figure 6** Comparison of the two principal coronal emission lines, forbidden Fe X and forbidden Fe XIV, in the western limb area.

At the beginning of a new solar cycle, the butterfly diagram records the appearance and the evolution of sunspots in medium heliographic latitudes (≈15° to ≈40°) above (or below) the solar Equator. On the other hand, during the end of the solar cycle sunspots appear close to the solar Equator [≤ |10°|]. Thus it seems that the heliographic distribution of Fe XIV probably follows a similar butterfly diagram. This correlation is expected as the region immediately above sunspots exhibits high activity (eruptive prominences, flares, *etc*.), associated with higher temperatures.

After a thorough mathematical analysis, the weak forbidden coronal lines Ni XIII (5115.8 Å – 350 eV – $T_{ionization}$ = (1.2 – 2.8)×$10^6$ K), Ar X (5539.1 Å – 423 eV – $T_{ionization}$ = (0.35–3.0)×$10^6$ K)) and the extremely weak line Ca XV (5694 Å – 820 eV – $T_{ionization}$ = (2.2 – 7.0)×$10^6$ K)) (Jefferies *et al.*, 1970; Gibson 1973; Mazzotta *et al.*, 1998) became visible (Figure 5a).

Forbidden Ca XV emission needs the exceptionally high coronal temperatures that often occur during solar maximum and, in particular, in energetic regions, such as coronal regions above solar flares. However, according to the work of Mazzotta *et al.* (1998), emission from multiple ionized elements (such as forbidden Ca XV) can happen even at lower temperatures, albeit with an extremely low intensity, which makes the detection of the emission line very difficult.

During the analysis and after the subtraction of the continuum background from the composite flash spectrum, a faint forbidden emission line of Ca XV (5694 Å) was observed (Figure 5a). Despite the fact that this line is extremely weak (200 times fainter than the Fe XIV line, Figure 5c), it follows the circular solar limb pattern, similar to the stronger emission lines. As can be readily seen from the width of the line, the spatial distribution of [Ca XV] (5694 Å) is extended, signifying that the contribution originates from a wide coronal area covering a large range of heights on the solar corona.

As mentioned in the previous paragraph, the forbidden Ca XV (5694 Å) emission line has been detected previously. However, the forbidden Ca XV (5446 Å) line was not detected in our data. This line is known to be very weak. Its ratio ($I_{5694Å}$ /$I_{5446Å}$) is not constant, varying between 1.67 and 3.13 (Chevalier and Lambert, 1970). During the total solar eclipse of 30 May 1965, the 5694 Å line was detected, but the 5446 Å line was undetectable (Curtis, Dunn, and Orrall, 1965). Furthermore, according to BASS2000 (see also Kubát *et al.*, 2010) any telluric absorption lines that could exist in the spectral band around the forbidden Ca XV emission line should be extremely weak, well below the sensitivity of our instrument.

4.3. The Height of the Chromosphere Deduced from the Mg I, Fe I / II, Na I, Hα and He I ($D_3$) Emission Lines

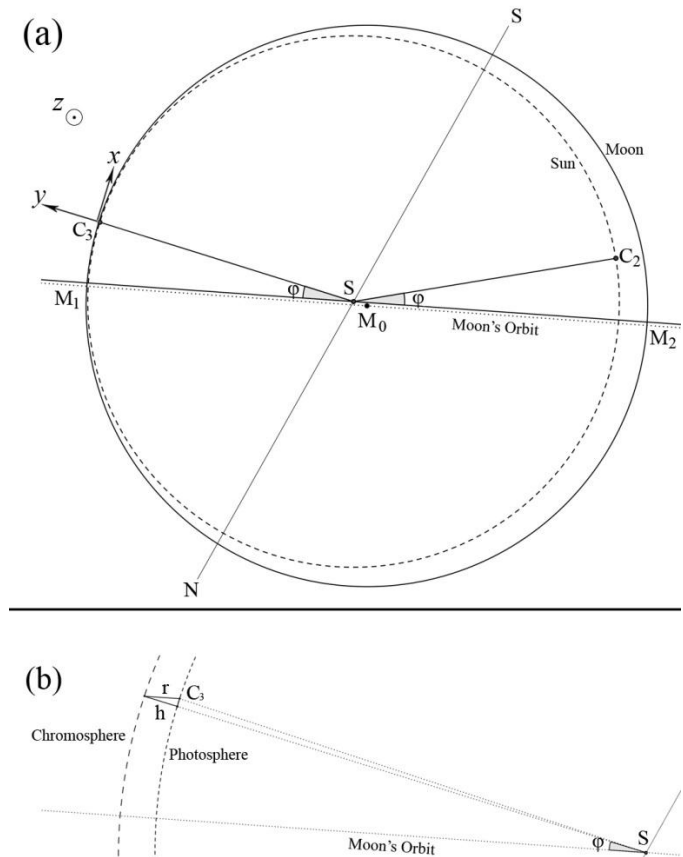

The emission lines of the light elements H and He were visible many seconds after the second contact and before the third contact (see also Voulgaris *et al.*, 2010). This persistence reflects the fact that these elements are distributed over large heights above the photosphere. This is corroborated by the visibility of prominences at large heights during the middle of totality. On the contrary, the emission lines of the elements heavier than He, such as Fe, Mg and Na, became visible very briefly (a few seconds only) after the second contact (and before the third contact), because these elements are distributed only in lower heights. It should be noted that during the middle of the eclipse of 11 July 2010, no emission lines of these elements were observed, because the region where they are formed was entirely covered by the lunar disk. The ratio of the angular diameter of the Moon to the angular diameter of the Sun was 1.05646 (Espenak and Anderson, 2008; Jubier, 2010). This relatively large diameter means that during totality, the Moon was covering a large portion of the Sun's atmosphere well above the photosphere. During the middle of totality, the Moon was covering an extra 39290 km above the eastern and western limbs. This extra coverage during the totality is sufficient to completely cover the chromosphere.

**Figure 7** a) Top: Second ($C_2$) and third ($C_3$) contacts with respect to the plane of the Moon's orbit. The orientation of the system agrees with Figures 3 and 5. b) Bottom: Calculation of height *h* of the chromosphere as a function of length *r*. The segment *r* is parallel to the moon's orbit.

At our observing location at Easter Island, we were 26.6 km off the center line of the total solar eclipse, so the eclipse was not central from our observing place. During totality, the projected movement of the center of the Moon ($M_1M_0M_2$), with respect to the center of the Sun (S), was in a straight line at constant speed. According to Jubier (2010), the angle $M_1SC_3$ is φ = 13.25° (Figure 7a) and the point of the third contact took place at the polar angle 275°. (With a smooth lunar limb profile, C3 would be at 282°, but because of a lunar valley located roughly between 270° and 280°, the real C3 is shifted).

During totality the integrated intensity $[\tilde{I}_E(\lambda, t)]$ of the distinct emission lines of each chromospheric element, in a suitable region $[\lambda - \Delta\lambda/2, \lambda + \Delta\lambda/2] \times [\varphi_1, \varphi_2]$ was measured as a function of time, $t$. $\Delta\lambda$ was chosen to be twice the width of the H$\alpha$ line (50 Å). The calculation of the integrated intensity was performed by the relation

$$\tilde{I}_E(\lambda, t) = \int_{\lambda - \Delta\lambda/2}^{\lambda + \Delta\lambda/2} f_t(\lambda) \, d\lambda \tag{9}$$

where $f_t(\lambda)$ is the function of the integrated intensity of emission lines along the circumference of the images of the solar disk, (eq. 8), from polar angle $\varphi_1 = 270°$ to $\varphi_2 = 280°$, as a function of time $t$. In this interval of polar angles, no prominences were present. The integrated intensity $[\tilde{I}_E(\lambda, t)]$ is directly proportional (for suitably small $\Delta\varphi$) to the customary integrated intensity $[E(\lambda, t)]$ that was introduced by Cillié and Menzel (1935) and used by Dunn (1968):

$$E(\lambda, t) = \int_{-\infty}^{\infty} \int_{h(t)}^{\infty} \int_{-1/2}^{1/2} \varepsilon_\lambda(x, y, z) e^{-\tau_\lambda} \, dx \, dy \, dz \tag{10}$$

where $\varepsilon_\lambda(x, y, z)$ is the volume emission coefficient at a point in the chromosphere and $\tau_\lambda$ is the optical depth measured along the line of sight from that point to the observer. The rectangular coordinate system is shown in Figure 7a (the $z$-axis is normal to the $xy$-plane toward the observer) and $h(t)$ is the height of the Moon's limb above the photosphere (transit). When $h = 0$, the intensity of the emission lines of all elements reaches its maximum value. This maximum occurs at second and third contacts (Voulgaris *et al.*, 2010). We define the maximum height that each element $[i]$ reaches, as the height $[h_{i,\max}]$ above the photosphere where the emission falls below our detection limit (Figure 8). If the eclipse is central, the two contacts points $[C_2$ and $C_3]$ are diametrically opposite and the Moon's motion is perpendicular to the tangent of the solar disk at the contact points along the height $h$ of the chromosphere. Since our observing site was not directly on the central line, the direction of the motion of the Moon $[r]$ was at an angle $\varphi = 13.25°$ with respect to the height $[h]$ above the contact points (Figure 7b). We can easily calculate the length $r$ for any element, through the relation $r = v\Delta t$, where $\Delta t$ is the length of time during which the emission line of the element was detectable, and $v$ is the velocity of projected cover of the solar disk by the Moon. In our case, $v$ was 285 km s$^{-1}$ (Jubier, 2010). The height $[h]$ is calculated from $h \approx r \cos\varphi = 0.97\, r$ (Figure 7b).

The graphs of the integrated intensity $[\tilde{I}_E(\lambda, t)]$ of various elements as a function of time and height are presented in Figure 8. It is obvious from this figure and from Table 1 that the height of the chromosphere measured by the integrated intensity of the emission lines of the light elements H and He is greater than or equal to $h_1 = 9400$ km (equivalent duration 34 seconds). In addition, the height of the lower chromosphere from the emission lines of the heavier elements Na, Mg, and Fe is greater than or equal to $h_2 = 3300$ km (equivalent duration 12 seconds), *i.e.*, $h_1/h_2 = 2.85$. We note that the equivalent duration of the Mg line during the 2008 eclipse was 8.65 seconds, which corresponds to 3290 km (Voulgaris *et al.*, 2010, Section 2.3.2).

For central-line total eclipses and assuming that the height of the chromosphere is 9400 km (Table 1) and if the ratio of the angular diameter of the Moon $(\theta_☾)$ to the angular diameter of Spectroscopic Coronal Observations during the Total Solar EclipseSpectroscopic Coronal Observations during the Total Solar Eclipse the chromosphere is $\geq 1$ (*i.e.* when the ratio $\theta_\mathrm{☾}/\theta_\odot \geq 1.0134$), then, during the middle of totality, no chromospheric emission lines of H and He are detected in the flash spectrum (with the exception of large prominences). For $1.0 \leq \theta_\mathrm{☾}/\theta_\odot \leq 1.0047$ the emission lines of heavier elements of the lower chromosphere (Table 1) are visible in the middle of totality (this mainly concerns hybrid eclipses). We note that the ratio $\theta_\mathrm{☾}/\theta_\odot$ of the 11 July 2010 eclipse was 1.05646; therefore, it is not surprising that we have not detected emission lines from Hα and He during the middle of totality (Figure 8).

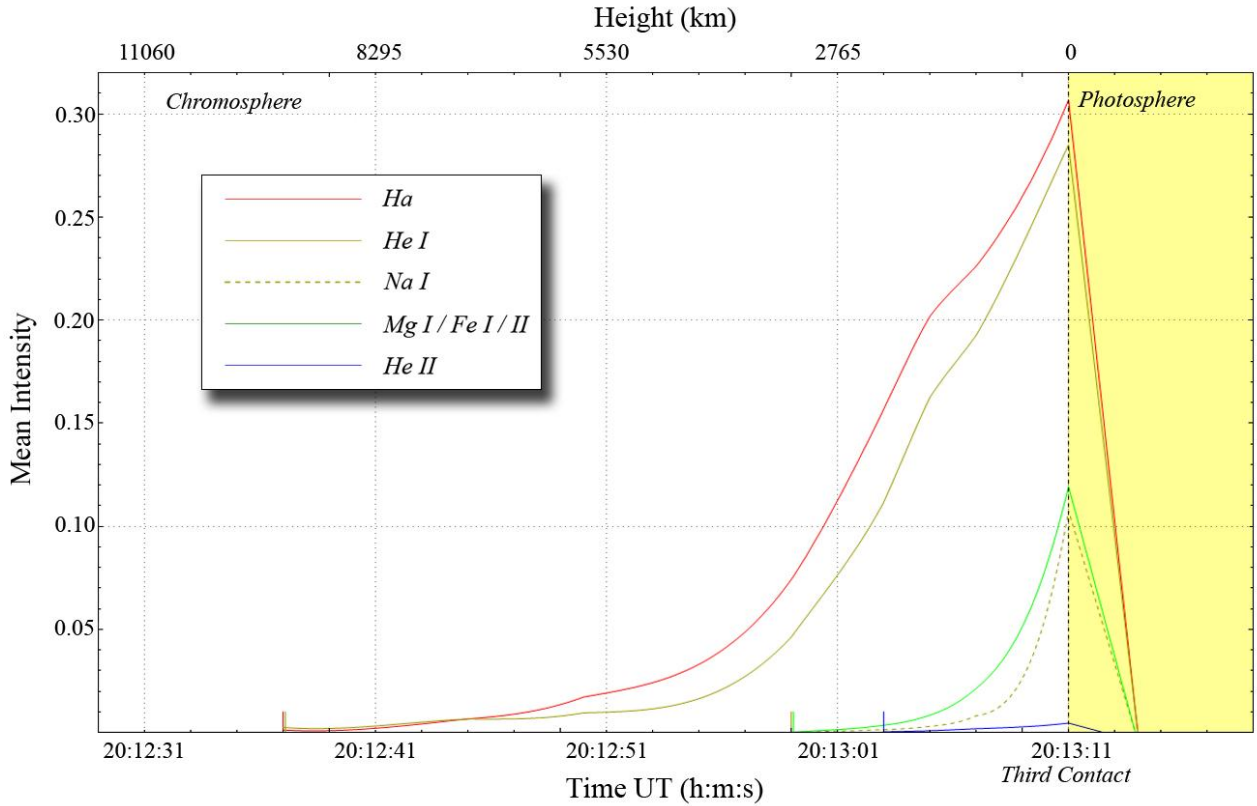

**Figure 8** The intensity of the emission lines of Hα (6563 Å), He I (5876 Å), Na I (5896, 5890 Å), Mg I / Fe I /II (5184, 5173, 5167 Å), and He II (4686 Å) as function of time (bottom) and height above the photosphere (top). The maximum intensity of each emission line occurs at third contact.

**Table 1** The duration of emission and the minimum height above the photosphere where the basic chromospheric elements can be detected

| Emission line | Duration of emission (s) | Minimum height (km) |
|---|---|---|
| Hα | 34 | 9400 |
| He I | 34 | 9400 |
| Na I | 12 | 3300 |
| Mg I/ Fe I / II | 12 | 3300 |
| He II | 8 | 2200 |

1313

4.4. The Emission Line of He II (4686 Å) in the Transition Zone

In our flash spectra, an exceptionally weak forbidden emission line of ionized helium at 4686 Å – 4685.7 Å and 4685.4 Å blended lines (NIST) – was visible during third contact. This line is exceptionally strong in some variable stars, *e.g.,* in the irregular type-1 variable star V617

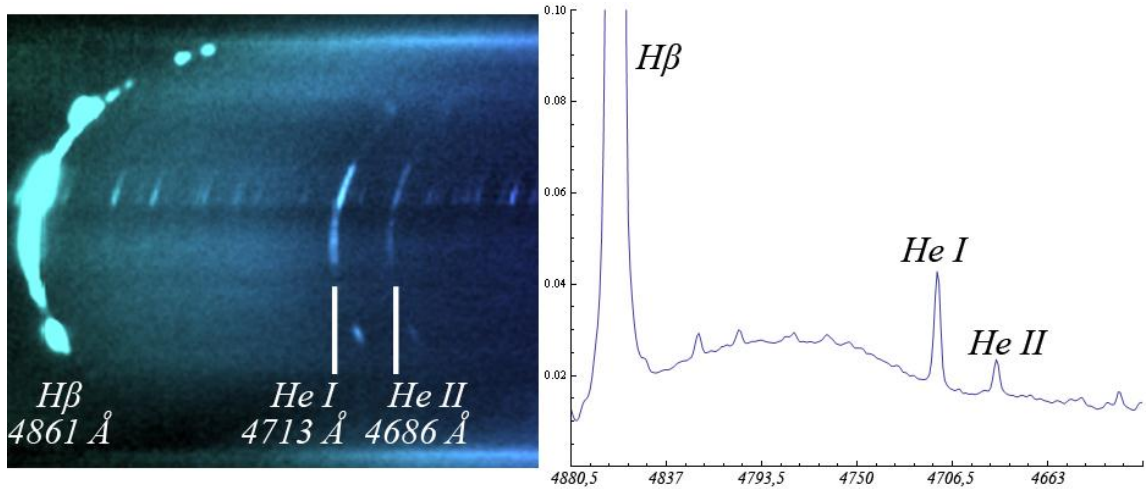

**Figure 9** (Left) An enlargement of part of Figure 3d. The emission lines of elements Hβ, He I, and He II are visible in the flash spectrum. (Right) The integrated photometric profile along the white line on either side of Figure 3a.

Sgr (Cieslinski, Diaz, and Steiner, 1999; Steiner *et al*., 1999). In these stars, the detected forbidden He II emission line is two–three times stronger than the Hβ line. In the data presented in Figure 9, the intensity of the forbidden He II line is 150 times weaker than the Hβ line (4861 Å) and four times weaker than the He I 4713 Å line.

The ionization energy of He II is 54 eV, and the ionization temperature is about 50000 K (Mariska, 1992). The He II layer is distributed in the transition zone (Bazin *et al.,* 2011) between the chromosphere and the corona, where there is a rapid increase of the temperature. In this zone several low ionization elements, like C IV, Si IV, O IV, N IV, have been detected in UV, from space telescopes (Sandlin *et al*., 1977; Dere *et al*., 1984). The detection of these lines is difficult in visible light, since the majority of these elements radiate mainly in the UV part of the spectrum. From our He II data the transition zone extends up to 2200 ± 280 km, indicating that this layer is very thin. The FWHM of this line has been measured to be 0.25 Å (Hirayama and Makoto, 1984).

Non-detections include the forbidden emission line of He II at 5411.5 Å, as well as forbidden emission lines of several other transition-zone elements, such as C IV (4658.3 Å, 5801.33 Å, 5811.98 Å) and Si IV (4088.85 Å, 4116.1 Å, 4631.24 Å, 4654.32 Å) (NIST). Of course, the permitted He II Lyman alpha line at 304 Å is well mapped from space.

**5. Conclusions**

In this work, we have described the analysis of the spectral data obtained during the total solar eclipse of 11 July 2011, at Easter Island, Chile. This method can be used in the analysis of any flash-spectrum data. The Easter Island eclipse coincided with the start of the new cycle, Solar Cycle 24, which commenced with relatively increased activity. The process of the subtraction of the continuum background from all spectra

*(i)*     revealed the extended distribution of the basic lines emitted by the solar corona, and
*(ii)*    enhanced the exceptional weak forbidden emission coronal lines Ni XIII, Ar X, and Ca XV.

In spite of the fact that forbidden Ca XV is usually emitted in regions with very high temperature, we detected it in regions of relatively low temperature. Besides the background subtraction, we transformed our spectra to a linear scale of wavelengths and converted the heliographic circular solar emission lines into linear topographic projections, which helped us recognize several previously unknown emission lines and integrate their intensity.

By comparing the distribution of the forbidden Fe X and Fe XIV emission lines, it is obvious that the regions from which they are emitted do not coincide, as has been shown by Takeda *et al*. (2000) from eclipse observations, by Singh *et al*. (2002, 2003) from coronagraph observations, and by others. Forbidden Fe XIV is detected at large coronal heights whereas forbidden Fe X is detected closer to the chromosphere. The heliographic distribution of forbidden Fe XIV probably follows a pattern similar to the well-known butterfly diagram. The height of the chromosphere was measured using the method of lunar disk transit on the solar disk using the light elements H and He, whereas the height of the lower chromosphere was measured using heavier elements such as Na, Mg and Fe. The height of the transition zone was determined by the same method using the high-temperature ion He II that is so well mapped with the *Extreme-ultraviolet Imaging Telescope* on the *Solar and Heliospheric Observatory* and with the *Atmospheric Imaging Array* on the *Solar Dynamics Observatory*. It will be interesting to compare our current results from the Easter Island total solar eclipse at the beginning of the solar maximum with the results we expect to obtain during the next total solar eclipse, in Australia on 14 November 2012 (13 November 2012 UT), which will happen much closer to the solar maximum.

**Acknowledgements**

We thank the Research Committee of the Aristotle University of Thessaloniki for financial support, E. Vanidhis for valuable help with optical laboratory measurements, Dimitrios Tsampouras of the Planetarium of Thessaloniki and George Pistikoudis and Spyros Kanouras for their assistance. We thank Robert Lucas, Muzhou Lu, and Craig Malamut for assistance on site.  JMP's current eclipse research is supported in part by the Solar Terrestrial Research program of the Astrophysics and Geospace Sciences Division of the National Science Foundation through grant AGS-1047726.  The 2010 expedition received support from the Brandi Fund and Science Center funds from Williams College.  JMP's solar research during the period of this eclipse and analysis has also been supported in part by grant NNX10AK47A from NASA's Marshall Space Flight Center. At the University of Chicago, partial financial support was provided by the Enrico Fermi Institute's Research Fund EFI 2-60190 to TEE.